\documentclass[prd,nofootinbib,preprint,superscriptaddress,twocolumn,10pt]{revtex4}
\pdfoutput=1
\usepackage[T1]{fontenc}
\usepackage{amsmath,amssymb}
\usepackage{braket}
\usepackage{epsfig}
\usepackage{graphicx}
\usepackage[usenames,dvipsnames]{color}
\usepackage{subfigure}
\usepackage{slashed}
\usepackage[colorlinks,citecolor=blue]{hyperref}
\usepackage{pdfpages}
\usepackage[normalem]{ulem}
\usepackage{color}
\usepackage{comment}
\usepackage{tikz-feynman}

\begin{document}


\title{Possible origin of the KM3-230213A neutrino event from dark matter decay}

\author{Debasish Borah}
\email{dborah@iitg.ac.in}
\affiliation{Department of Physics, Indian Institute of Technology Guwahati, Assam 781039, India}
\affiliation{Pittsburgh Particle Physics, Astrophysics, and Cosmology Center, Department of Physics and Astronomy, University of Pittsburgh, Pittsburgh, PA 15260, USA}

\author{Nayan Das}
\email{nayan.das@iitg.ac.in}
\affiliation{Department of Physics, Indian Institute of Technology Guwahati, Assam 781039, India}

\author{Nobuchika Okada}
\email{okadan@ua.edu}
\affiliation{Department of Physics, University of Alabama, Tuscaloosa, Alabama 35487, USA}

\author{Prantik Sarmah}
\email{prantiksarmah@ihep.ac.cn}
\affiliation{Institute of High Energy Physics,
Chinese Academy of Sciences, Beijing, 100049, People’s Republic of China}

\begin{abstract}
We study the possibility of the highest energy neutrino event with 220 PeV energy, detected recently by the KM3NeT experiment to be originating from heavy dark matter (DM) decay. Considering a heavy right handed neutrino (RHN) DM for illustrative purpose, we show that DM mass of 440 PeV, can explain the observed flux. The required DM lifetime to produce the best-fit value of the neutrino flux saturates the existing gamma-ray bounds. Due to the large uncertainty in the flux, it is possible to explain the KM3NeT event from RHN DM decay at $3\sigma$ confidence level (CL) while being in agreement with gamma-ray bounds and non-observation of similar events at IceCube. While we consider a gauged $B-L$ scenario where DM relic can be generated due to other interactions, we also briefly discuss some alternate DM possibilities where the gamma-ray bounds can be alleviated compared to the minimal RHN DM discussed here.
\end{abstract}
\maketitle
\section{Introduction}
Recently, the KM3NeT collaboration has reported the detection of a neutrino event named as KM3-230213A having $\mathcal{O}(100 \, \rm PeV)$ energy \cite{KM3NeT:2025npi}. The deep-sea neutrino telescope has detected an ultra-high-energy (UHE) muon with energy $120^{+110}_{-60}$ PeV arriving from a near-horizontal direction (RA: $94.3^{\circ}$, Dec: $-7.8^{\circ}$) which is most likely to be created by a neutrino of even higher energy in the vicinity of the detector. The required neutrino energy was found to be in the range 110-790 PeV at $1\sigma$ CL with median energy being 220 PeV. This is the highest energy neutrino event ever detected. Creation of such energetic neutrinos through cosmic ray interactions ($pp$ or $p\gamma$) in conventional astrophysical sources would require protons to be accelerated to around EeV energies. However, such sources are not available in our Milky Way or any nearby Galaxies. The KM3NeT collaboration has analyzed several classes of source candidates such as blazars, gamma-ray bursts using different electromagnetic telescope catalogs and found no conclusive evidence~\cite{KM3NeT:2025npi}. The absence of any conclusive astrophysical source \cite{KM3NeT:2025bxl, KM3NeT:2025aps} points towards a cosmic origin of the highest energy neutrino event detected so far \cite{KM3NeT:2025vut}. Cosmogenic UHE neutrinos together with UHE gamma-rays are expected to be created in the interactions of UHE cosmic rays with the cosmic microwave background (CMB) photons, the so called Greisen–Zatsepin–Kuzmin (GZK) process~\cite{PhysRevLett.16.748,1966JETPL...4...78Z}. Several model estimates of these UHE neutrinos and gamma-rays flux can be found in the literature~\cite[e.g., see ][]{Gelmini:2005wu,Aloisio:2015ega,Gelmini:2022evy,Chakraborty:2023hxp}. The UHE neutrino flux required to produce the  KM3-230213A event was found to overshoot not only these model estimates, but also the sensitivities of other presently running neutrino telescopes such as IceCube and Pierre Auger Observatory (PAO). This could be due to the large uncertainty (about 3 orders of magnitude) in the  KM3-230213A flux, possibly created by an upward fluctuation~\cite{KM3NeT:2025npi}.
On the other hand, non-observation of similar events at other experiments like IceCube and PAO leads to a $2.5\sigma-3\sigma$ tension\footnote{An independent analysis \cite{Li:2025tqf} finds a slightly larger tension.} with the cosmogenic origin hypothesis of the KM3-230213A event \cite{KM3NeT:2025ccp}. This tension is significant given the $1\sigma$ CL estimate of the neutrino energy being above 100 PeV. Several follow-up studies on implications of this neutrino event for UHE cosmic rays \cite{Muzio:2025gbr, Fang:2025nzg}, Lorentz invariance violation \cite{KM3NeT:2025mfl,Satunin:2025uui,Yang:2025kfr}, possible astrophysical origin \cite{Dzhatdoev:2025sdi, Neronov:2025jfj}, possible origin from primordial black hole evaporation \cite{Boccia:2025hpm}, in-vacuo dispersion of neutrinos \cite{Amelino-Camelia:2025lqn} have also appeared in the literature.

In this work, we consider a dark matter (DM) origin of the KM3-230213A event flux at $3\sigma$ CL while also addressing the the mild tension with IceCube and PAO for such diffuse, isotropic origin of the event. DM decay origin of the PeV neutrino events at IceCube \cite{IceCube:2013cdw, IceCube:2013low} more than a decade ago was also explored in several works including \cite{Murase:2015gea, Cohen:2016uyg, Sui:2018bbh, Dev:2016qbd, Borah:2017xgm}. Prospects of probing heavy DM of a much wider mass range at neutrino telescopes have been studied in \cite{Chianese:2021htv}. Corresponding bounds on such heavy DM lifetime from gamma-ray observations can be found in \cite{LHAASO:2022yxw, Chianese:2021jke}. While there is no particular advantage of the DM-origin hypothesis compared to the upward fluctuation \cite{KM3NeT:2025npi}, we consider it to be an appealing alternative offering an interesting detection aspect of particle DM. Given the uncertainty in the flux of the KM3-230213A event, our proposal for such a cosmogenic origin can still be consistent with non-observations at IceCube and PAO if the actual flux remains below the their respective sensitivities. As noted in \cite{KM3NeT:2025npi}, the $3\sigma$ CL of the KM3-230213A event flux is $(0.02-47.7) \times 10^{-8} \, {\rm GeV}\, {\rm cm}^{-2} \, {\rm s}^{-1} \, {\rm sr}^{-1}$ which extends beyond IceCube and PAO sensitivities by at least one order of magnitude. Another motivation behind this DM-origin hypothesis is the opportunity to constrain lifetime of superheavy DM close to EeV scale using UHE neutrino. Previously, UHE neutrino data from IceCube were used to constrain lifetime of superheavy DM with PeV scale masses \cite{Esmaili:2014rma}.

We consider a heavy right handed neutrino (RHN) to be the DM candidate whose late decay into neutrinos is responsible for the highest energy neutrino event mentioned above. While the neutrino event can, in principle, originate from DM annihilation as well, the flux will be proportional to DM density squared and hence suppressed for such superheavy DM. Such a RHN appears in type-I seesaw scenario which explains the origin of light neutrino masses. While two RHNs are sufficient to generate light neutrino mass and mixing, the third RHN can be DM if sufficiently long-lived due to feeble Yukawa coupling with the standard model (SM) leptons. Since RHN can decay into other SM particles due to the same Yukawa interactions, we also have secondary gamma-ray productions, tightly constrained from astrophysical observations. We show that the required flux and energy of the KM3-230213A neutrino event can be explained within the uncertainties from decay of a RHN DM having mass $M_{\rm DM} \sim 440$ PeV while being in agreement with the gamma-ray bounds. While the required relic of such superheavy DM can not be generated in the minimal type-I seesaw model, we consider a gauged $U(1)_{B-L}$ extension of the model where relic can be generated non-thermally. We also comment on possible alternative DM realisations in which the gamma-ray bounds can be weaker due to phase space suppression. Depending upon the specific realisation and DM mass, gauged $U(1)_{B-L}$ offers very interesting multi-messenger detection prospects at energy, intensity as well as cosmic frontiers \cite{Bose:2022obr, Okada:2017pgr, Barman:2025bir, Borah:2022byb, Abazajian:2019oqj}.

This paper is organised as follows. In section \ref{sec1}, we discuss our model followed by the details of the flux calculation in section \ref{sec2}. We discuss our results in section \ref{sec3} and briefly discuss some other DM realisations to alleviate gamma-ray bounds in section \ref{sec3a}. We finally conclude in section \ref{sec4}.

\section{The Model}
\label{sec1}
We consider heavy right handed neutrino DM in a gauged $B-L$ extension of the SM \cite{Davidson:1978pm, Mohapatra:1980qe, Marshak:1979fm, Masiero:1982fi, Mohapatra:1982xz, Buchmuller:1991ce}. In its minimal realisation, the SM particle content is extended with three RHNs $N_i\, (i=1,2,3)$ and one complex singlet scalar ($\Phi$) all of which are singlet under the SM gauge symmetry. The requirement of triangle anomaly cancellation in this minimal model requires each of the RHNs to have $B-L$ charge $-1$. The complex singlet scalar having $B-L$ charge 2 not only leads to spontaneous breaking of gauge symmetry but also generate RHN masses or the seesaw scale dynamically. The Lagrangian involving RHN can be written as
\begin{align}
 \mathcal{L}_{\rm fermion} &=  i \sum_{\kappa=1}^{3}\overline{N_{R_{\kappa}}} \slashed{D} N_{R_{\kappa}} -\bigg (\sum_{\substack{j=1~\\ \alpha=e, \mu, \tau}}^{3}Y_D^{\alpha j}~\overline{l_{L}^{\alpha}}\tilde{H} N_{R}^{j}\nonumber\\
 & + \sum_{i,j=1}^{3}Y_{N_{ij}}\Phi~\overline{N_{R_i}^{C}}N_{R_j}
+{\rm h.c.} \bigg ),\label{eq:Lferm}
\end{align}
where $l_L, H$ denote the SM Lepton and Higgs doublets, respectively. The covariant derivative for $N_{R \kappa}$ is defined as 
\begin{eqnarray}
\slashed{D}\,{N_{R_{\kappa}}} =
\gamma^{\mu}\left(\partial_{\mu}
- i g_{\rm BL}\,Z'_{\mu}\right) {N_{R_{\kappa}}} \,, 
\end{eqnarray}
with $g_{\rm BL}, Z'$ being the coupling and gauge boson respectively for $U(1)_{B-L}$. After the singlet scalar $\Phi$ acquires a non-zero vacuum expectation value (VEV) denoted by $v_{\rm BL}$, the RHNs and $Z'$ acquire masses as, 
\begin{equation}
 M_{Z'}=2 g_{\rm BL} v_{\rm BL}, \,\,\,\, M_{i}=\sqrt{2}Y_{N_{i}}v_{\rm BL},\label{eq:DMmass}
\end{equation}
where we have assumed the Yukawa coupling $Y_N$ to be diagonal. We consider the lightest RHN namely $N_1$ to be the DM candidate such that the corresponding Yukawa coupling $Y^{\alpha 1}_D$ to ensure a long lifetime. Due to such Yukawa interactions which also mixes $N_1$ with the light neutrinos after electroweak symmetry breaking, we can have dominant tree-level decays like $N_1 \rightarrow \nu, h; \nu, Z; l^{\mp}, W^{\pm}$ with $h$ being the SM-like Higgs. Ignoring the masses of SM particles in comparison to $N_1$ mass $M_1$, these decay widths are given as \cite{Higaki:2014dwa}
\begin{align}
    \Gamma (N_1 \rightarrow \nu_\alpha \,h (Z)) = \frac{\lvert Y^{\alpha 1}_D\rvert^2 M_1}{32 \pi},  \nonumber  \\
     \Gamma (N_1 \rightarrow l^-_\alpha \,W^+) = \frac{\lvert Y^{\alpha 1}_D\rvert^2 M_1}{16 \pi}.
     \label{decay}
\end{align}
We can also have a radiative decay $N_1\rightarrow \gamma \nu_\alpha$ as 
\cite{Pal:1981rm,Shrock:1982sc}
	\begin{eqnarray}
		\Gamma_{N_1\rightarrow \gamma \nu_\alpha}\simeq \frac{9G^2_{F}\alpha_{_{\rm EM}}}{1024\pi^4}\sin^2(\theta_{{\alpha 1}})M_{1}^5,
	\end{eqnarray}
    where $\alpha_{_{\rm EM}}=1/137$ is the fine structure constant, $G_F=1.166\times10^{-5}~\rm GeV^{-2}$ is the Fermi constant. The active-sterile mixing is given by $\sin(\theta_{{\alpha 1}}) \sim \frac{Y^{\alpha 1}_D v}{\sqrt{2} M_1}$ with $v$ being the VEV of the SM Higgs doublet. The lifetime of $N_1$, dictated by its tree-level two-body decays, can then be estimated as
\begin{equation}
    \tau_{N_1} \approx 8 \times 10^{28} \, {\rm s} \left ( \frac{440 \; {\rm PeV}}{M_1} \right ) \left ( \sum_\alpha \bigg \lvert \frac{Y^{\alpha 1}_D}{4.7 \times 10^{-31}} \bigg \rvert^2 \right)^{-1}
    \label{lifetimedm}
\end{equation}
which is identified as the DM lifetime $\tau_{\rm DM}$ in our setup. The dominant tree level decay widths given in Eq. \eqref{decay} have $2:1:1$ ratio for $l^{\pm}_L W^{\mp}, \nu Z, \nu h$ final states, respectively. Using the best-fit values of neutrino mixing angles for normal ordering (NO) \cite{Esteban:2024eli}, the branching ratio (BR) into different lepton flavours for a particular decay mode follows the ratio ${\rm BR}(e):{\rm BR}(\mu):{\rm BR}(\tau)=0.68:0.21:0.11$. The flavour ratio at source for inverted ordering (IO) turns out to be ${\rm BR}(e):{\rm BR}(\mu):{\rm BR}(\tau)=0.02:0.55:0.43$. We assume the Dirac CP phase to be zero for simplicity. The neutrino flavour ratio at the detector, after considering oscillation effects \cite{Athar:2000yw}, turns out to be approximately $3:2:1$ and $1:2:3$ for NO and IO respectively.

While DM of required mass can not be of thermal origin due to the violation of the unitarity bound \cite{Griest:1989wd}, it can not be produced non-thermally via Yukawa portal interactions as well due to the suppressed Yukawa couplings in Eq. \eqref{lifetimedm} required to ensure its longevity. However, due to additional interactions present in the $B-L$ model which are unrelated to DM lifetime, it is still possible to produce the desired relic via the freeze-in mechanism \cite{Hall:2009bx, Biswas:2016bfo, Borah:2020wyc, Okada:2020cue}. 

Using the approach of \cite{Biswas:2016bfo, Borah:2020wyc}, we consider the production of superheavy $N_1$ of mass 440 PeV from decay of heavier particle. The required non-thermal criteria forces the gauge coupling $g_{\rm BL}$ to be tiny keeping $Z'$ out-of-equilibrium as well. We consider the singlet scalar to be in equilibrium sourcing $Z'$ with the latter acting as source of DM. The relic can be calculated by solving the coupled Boltzmann equations written in terms of comoving number densities $Y_{X} = \frac{n_{X}}{s}$ as follows:
\begin{eqnarray}
    \frac{dY_{Z'}}{dx} = \frac{\beta}{ x \mathcal{H}}  \Gamma_{\phi \to Z' Z'} \frac{K_{1}(M_{\phi}/T)}{K_{2}(M_{\phi}/T)} Y^{\rm eq}_{\phi}  \nonumber\\ \nonumber - \frac{\beta}{ x \mathcal{H}}  \Gamma_{Z' \to \text{all}} \frac{K_{1}(M_{Z'}/T)}{K_{2}(M_{Z'}/T)} Y_{Z'},  \\
    \frac{dY_{N_{1}}}{dx} = \frac{\beta}{ x \mathcal{H}}  \Gamma_{\phi \to N_{1} N_{1}} \frac{K_{1}(M_{\phi}/T)}{K_{2}(M_{\phi}/T)} Y^{\rm eq}_{\phi}\nonumber \\ + \frac{\beta}{ x \mathcal{H}}  \Gamma_{Z' \to N_{1} N_{1}} \frac{K_{1}(M_{Z'}/T)}{K_{2}(M_{Z'}/T)} Y_{Z'}.
    \label{dmeq}
\end{eqnarray}
Here $x= M_{Z'}/T$, $\beta = 1 + \frac{T d g_{s}/dT}{3 g_{s}}$, with $g_{s}$ being the relativistic entropy degrees of freedom. $\mathcal{H}$ is the Hubble expansion parameter, $\Gamma$ denotes the decay width and $K_i$'s denote the modified Bessel functions of the second kind.

\section{Neutrino and Gamma-ray flux from DM decay}
\label{sec2}
In this section, we describe the method to compute the flux of high energy neutrinos and gamma-rays produced from DM decay. As discussed above, we consider the RHN $N_{1}$ decay via all possible two-body decay modes namely, $N_1 \rightarrow \nu, h (Z, \gamma)$ and $N_1 \rightarrow \ell^{\pm}, W^{\mp}$, emitting secondaries such as high energy neutrinos, gamma-rays and electrons. Note that we do not distinguish between particle and anti-particle final states here. The DM particle $N_1$ is considered to be the constituent of the Milky Way DM halo. Thus, the decay of $N_{1}$ in the Milky Way can give rise to backgrounds of high energy neutrinos and gamma-rays. The flux of these secondaries $\phi_{i, \rm G}(E_i)$ from Galactic (G) DM can be estimated as follows~\cite{Murase:2012xs,Sarmah:2024ffy}:
\begin{equation}
    \frac{\mathbf{d}^2 \phi_{i, \rm G} (E_{i})}{\mathrm{d} E_{i} \mathrm{d} \Omega} = \frac{\mathcal{D}}{4 \pi M_{\rm DM} \tau_{\rm DM} }  \frac{\mathrm{d} N_{i}(E_i)}{\mathrm{d}E_{i}}~,
        \label{eq:Flux-formula1}
\end{equation}
where, $E_{i}$ denotes the energy of the $i$th ($\nu$ or $\gamma$) particle and $M_{\rm DM} \equiv M_1$ is DM mass with lifetime denoted by $\tau_{\rm DM}$. The dependence of the flux on DM density profile is given by~\cite{Maity:2021umk},
\begin{equation}
    \mathcal{D}  = \frac{1}{\Delta \Omega} \int_{\Delta \Omega} \mathrm{d} \Omega \int_{0}^{s_{max}} \mathrm{d}s~ \rho_{\rm DM} (s, b, l). 
    \label{eq:Flux-formula}
\end{equation}
Here, $\Delta \Omega$ is the angular region of observation, $(l,b)$ are Galactic coordinates (longitude and latitude), and $s$ is the line-of-sight coordinate. These coordinates are related as $r (s,b,l) = \sqrt{s^2 + r_{\odot}^2- 2 s r_{\odot} \cos{b} \cos{l}}$. The differential solid angle, $\mathrm{d} \Omega$ is given  by $\mathrm{d} \Omega = (\sin{b})\mathrm{d}l$ and $r_{\odot} = 8.3~\rm kpc$ is the distance to Milky Way centre from the Sun. The DM density, $\rho_{\rm DM} (s,b,l)$  can be obtained from various numerical models in the literature~\cite{Navarro:1996gj,Navarro:2003ew,1989A&A...223...89E,Graham:2005xx,Burkert1995}. For this work, we consider the Navarro-Frenk-White (NFW) density profile~\cite{Navarro:1996gj} as a function of the Galactocentric radius, $R_{\rm GC}$ given by,
\begin{equation}
    \rho_{\rm NFW}(R_{\rm GC}) = \frac{\rho_{\rm c}}{(R_{\rm GC}/R_{\rm c})(1+R_{\rm GC}/R_{\rm c})^2} \ ,
\end{equation}
 where $\rho_{\rm c}$ is the  DM density at the characteristic scale $R_{\rm c}=11$~kpc, and its value is obtained by normalizing $\rho_{\rm NFW}$ with the DM density in the solar neighbourhood, $\rho_{\odot}=0.43~\rm GeV~cm^{-3}$. Note that all the  DM profiles in the literature nearly predict similar DM density except at the Galactic center. Therefore, the variation of the DM profile is only important for signals from the Galactic center direction. Since the arrival direction of the KM3-230213A event is far away from the Galactic center direction, we do not consider other DM profiles in this work.

In addition to the Galactic contribution, the DM decay in the Extra-galactic (EG) space can also produce such secondaries via the same decay modes. While the  contribution  of the EG DM  to gamma-ray flux can be expected to be negligible due to the absorption in the extra-galactic background light (EBL) and the CMB, the  contribution to the neutrino flux can be important. This is because neutrinos are not affected by absorption during propagation. The differential flux of the EG secondaries is given by,  
\begin{eqnarray} \label{eq:eg}
    \frac{d \phi_{i, \rm EG} (E)}{dE} = \frac{c\, \rho_{\rm DM}}{4\pi\, M_{\rm DM }  \tau_{\rm DM}} \int dz \left|\frac{dt}{dz}\right| \frac{dN (E')}{dE'} e^{-\tau (E^{\prime},z)}.\nonumber \\
\end{eqnarray}
Here, $E' = E(1+z)$ where $z$ is the red-shift factor and $\rho_{\rm DM} = \Omega_{\rm DM} \rho_{c}$, with $\rho_{c}=4.7\times 10^{-6}$ GeV cm$^{-3}$ denoting the critical density in a flat Friedmann–Lema${\rm \hat{i}}$tre–Robertson–Walker (FLRW) Universe and $\Omega_{\rm DM}=0.27$. The cosmological line element, $ \left|\frac{dt}{dz}\right|$ can be expressed as 
\begin{eqnarray}\label{eq:eg1}
     \left|\frac{dt}{dz}\right| = \frac{1}{\mathcal{H}_{0}(1+z)\sqrt{(1+z)^3 \Omega_{\rm m} + \Omega_{\Lambda}}},
\end{eqnarray}
where $\Omega_{m}=0.315$, $\Omega_{\Lambda}=0.685$ and $\mathcal{H}_{0} = 67.3$ Km s$^{-1}$ Mpc$^{-1}$ is the current expansion rate of the Universe. The factor $e^{-\tau (E^{\prime},z)}$ takes into account of the gamma-ray absorption during propagation, where $\tau (E^{\prime},z)$ is the total optical depth of EBL and CMB. The optical depth of EBL is taken from ~\cite{Stecker:2016fsg}.

The differential energy spectra, $  \mathrm{d} N_{i}/ \mathrm{d} E_{i}$ in Eq.~\eqref{eq:Flux-formula1} and Eq.~\eqref{eq:eg} are obtained from the publicly available code \texttt{HDMSpectra}~\cite{Bauer:2020jay}.  These spectra depend on the DM mass and the inclusive decay rate of the DM particle. The total flux of secondary gamma-rays and neutrinos is given by $\phi_{i} = \phi_{i,\rm G}+\phi_{i,\rm EG}$.

\begin{figure*}
    \centering
        \includegraphics[width=0.49\linewidth]{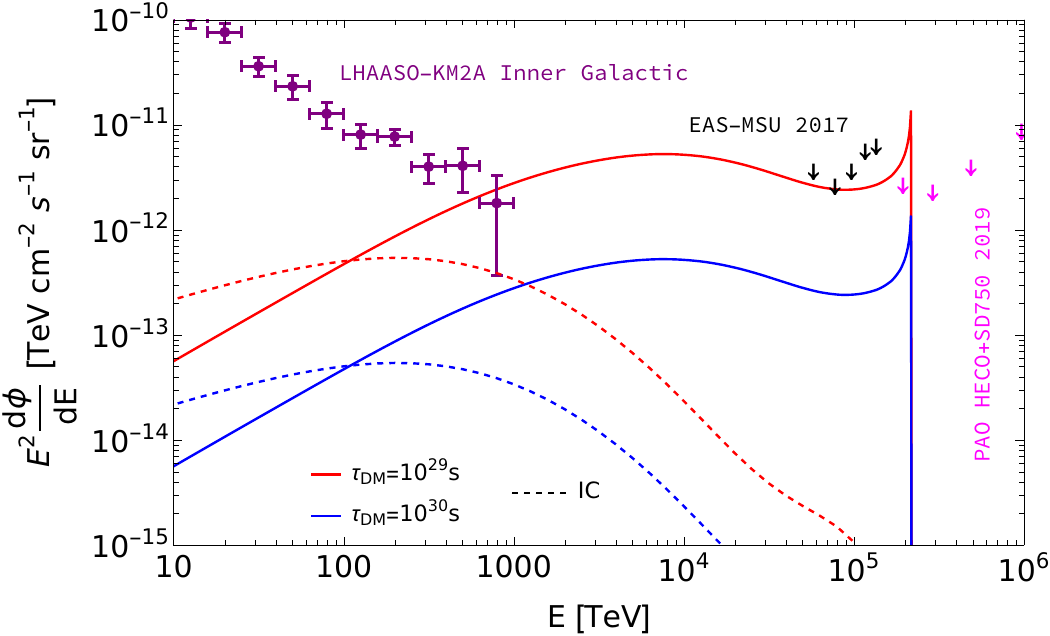}    
        \includegraphics[width=0.49\linewidth]{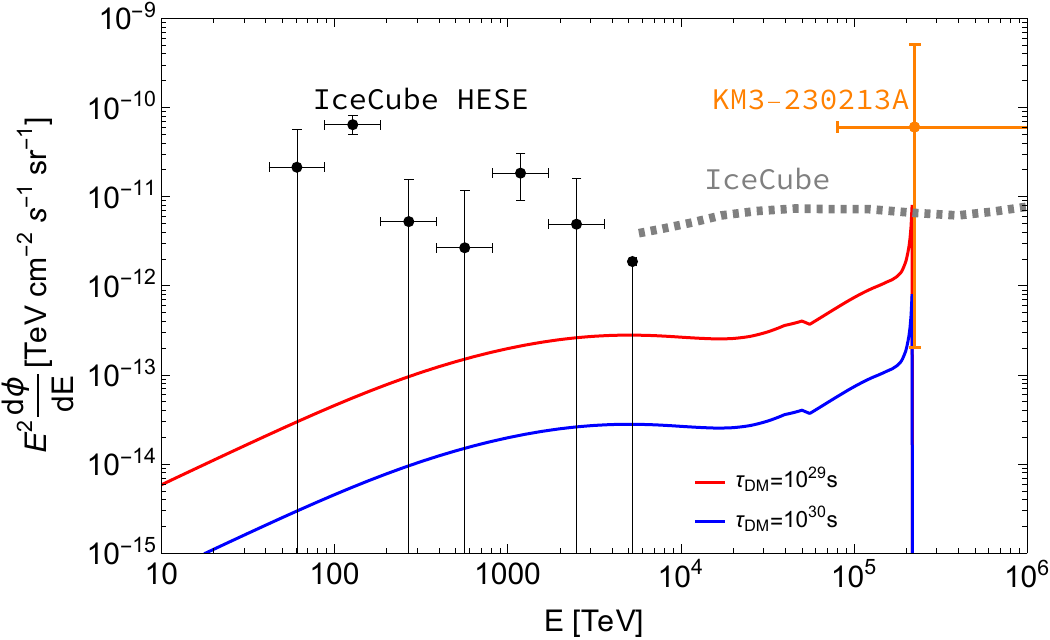}
    \caption{Gamma-ray (left-panel) and muon neutrino (right-panel) fluxes  with DM mass $M_{1}= 440~\rm PeV$ and two different values of DM lifetime $\tau_{\rm DM}=10^{29}$ s (solid red), $\tau_{\rm DM}=10^{30}$ s (solid blue). The dashed lines on the left-panel correspond to inverse Compton radiation. The fluxes are summed over the all possible decay modes and both Galactic and Extra-galactic contributions.  The data points on the left panel plot correspond to gamma-ray bounds from LHAASO \cite{LHAASO:2023gne}, EAS-MSU \cite{Fomin:2017ypo} and PAO~\cite{Castellina:2019huz}. The data points on the right panel plot correspond to UHE neutrino events at IceCube \cite{IceCube:2020wum} and KM3NeT \cite{KM3NeT:2025npi}. The gray dotted contour (right panel) denotes IceCube's sensitivity to UHE neutrino flux\cite{IceCube:2018fhm}.}
    \label{fig1}
\end{figure*}


\section{Results and Discussion}
\label{sec3}
In this section, we discuss the results obtained by computing the fluxes of UHE neutrinos and gamma-rays from DM decay using Eq.~\eqref{eq:Flux-formula1} and analyze them in view of the KM3-230213A event. The mass of the DM particle $N_1$ is choosen to be $440$~PeV in order to match the energy requirement ($220$~PeV) of the KM3-230213A event.  For this analysis, we take into account the constraints from three different gamma-ray observations. For  energies below $1$ PeV, we take the diffuse gamma-ray measurements from the inner Galactic plane by the Large High Altitude Air Shower Observatory (LHAASO)~\cite{LHAASO:2023gne}, whereas for energies above $40$ PeV, we consider the upper limits on UHE photon flux from Moscow State University Extensive Air Shower (EAS-MSU) array~\cite{Fomin:2017ypo} and PAO (PAO HECO +SD750)~\cite{Castellina:2019huz}. In addition to these gamma-ray constraints, we also consider the IceCube's  High Energy Starting Events (HESE)~\cite{IceCube:2020wum}. 

We  show the total   gamma-ray (left) and muon neutrino (right) fluxes together with these constraints in Fig.~\ref{fig1}. The fluxes are summed over all possible decay modes, and both Galactic and EG component. The gamma-ray fluxes shown here are computed for the inner Galactic plane ($15^{\circ} < l < 125^{\circ}, -5^{\circ} < b < 5^{\circ}$) to remain consistent with the LHAASO data (shown by the purple data points). This flux is slightly larger (about a factor of $\sim 1.25$) than the overall diffuse flux of the DM halo. For comparison, we also include the high energy diffuse gamma-ray limits, ESU-MSU and PAO HECO +SD750 shown by the black and magenta downward arrows, respectively. Given these constraints, the gamma-ray flux can be appropriately normalized by tuning the DM lifetime, $\tau_{\rm DM}$ (see Eq.~\eqref{eq:Flux-formula1}). To demonstrate this, the fluxes for two different values of $\tau_{\rm DM}$ are plotted in the left panel of Fig.~\ref{fig1}.  The flux  corresponding to $\tau_{\rm DM} \approx 10^{29}~s$  (red curve) remain consistent with the LHAASO data, however, it is in tension with the ESU-MSU and PAO HECO+SD750 limits. This tension can be relaxed by choosing a larger lifetime, $\tau_{\rm DM} \approx 10^{30}~s$ as shown by the blue curve.  Note that both Galactic and EG  gamma-ray fluxes can undergo absorption due to pair production losses on low energy background photons such as interstellar radiation field (ISRF), EBL and CMB. This absorption is negligible above $10$ PeV~\cite{Esmaili:2015xpa} for the Galactic gamma-ray flux. However, the absorption of the EG gamma-ray flux in the EBL and CMB is found to be severe. Thus, the total gamma-ray flux is completely dominated by the Galactic component.
Apart from absorption, the Galactic flux can also be affected by inverse Compton radiation from secondary electrons and positrons. We calculate the inverse Compton radiation for two specific values of $\tau_{\rm DM}$ using the method outlined in \cite{Esmaili:2015xpa}, and show the results as dashed curves with the same color scheme. For the target photons, we consider both CMB and ISRF. However, the contribution of ISRF is found to be negligible in the energy range of interest.  In computing the inverse Compton flux, we also account for the synchrotron energy losses of secondary electrons and positrons in the intergalactic magnetic field, assuming an average field strength of $1 ~\mu \rm G$. Since the synchrotron power is proportional to the square of the electron energy, these losses become severe at higher energies.   This, in turn, yields a softer inverse Compton flux at higher energies. Thus, in the energy range of our interest, the gamma-ray flux is solely dominated  by the secondary photons from DM decay.


For the two specific values of $\tau_{\rm DM}$, we show the corresponding muon neutrino fluxes in the right panel of Fig. \ref{fig1} following the same color coding as the left panel. The muon neutrino flux is computed for the  NO scenario considering the flavour ratio $3:2:1$ at Earth. The IO scenario having a flavour ratio $1:2:3$ produces similar neutrino fluxes and are not shown here. The EG component is found to contribute slightly to the total muon neutrino flux below  $50$~PeV. We also show the flux of the  KM3-230213A event by orange data point with $3\sigma$ CL. The black data points and the gray dotted curve, respectively, show the IceCube's HESE events~\cite{IceCube:2020wum} and UHE neutrino flux sensitivity at  $90\%$ CL~\cite{IceCube:2018fhm}. For the  estimation of the neutrino fluxes, we consider the arrival direction (RA: $94.3^{\circ}$, Dec: $-7.8^{\circ}$) of KM3-230213A with an angular uncertainty of $3^{\circ}$ as reported by the KM3NeT collaboration~\cite{KM3NeT:2025npi}. The flux obtained with this conservative approach is about a factor of $2$ smaller than the overall diffuse neutrino flux of the DM halo. Clearly, RHN DM with lifetime $\tau_{\rm DM}=10^{29}$ s and mass $M_1=440$ PeV can give rise to a flux close to the central value of the KM3-230213A event at 220 PeV energy. However, it also produces a large flux of secondary gamma-rays saturating the existing upper limits from LHAASO while being in tension with EAS-MSU and PAO data as discussed above. Lowering the decay width of DM or increasing the lifetime leads to agreement with gamma-ray bounds at the cost of reducing the neutrino flux while still being consistent at $3\sigma$ CL, as depicted by the flux corresponding to $\tau_{\rm DM} \approx 10^{30}~s$. Interestingly, this neutrino flux  can also alleviate the tension between KM3NeT and IceCube regarding the diffuse, isotropic origin of the KM3-230213A event. This tension arises as the central value of the  KM3-230213A flux is larger than the IceCube's sensitivity. While the present DM model can not simultaneously explain all UHE neutrino events including IceCube HESE and KM3-230213A, the former can be explained with astrophysical sources \cite{IceCube:2020wum}.

\begin{figure}
    \centering
    \includegraphics[width=1.0\linewidth]{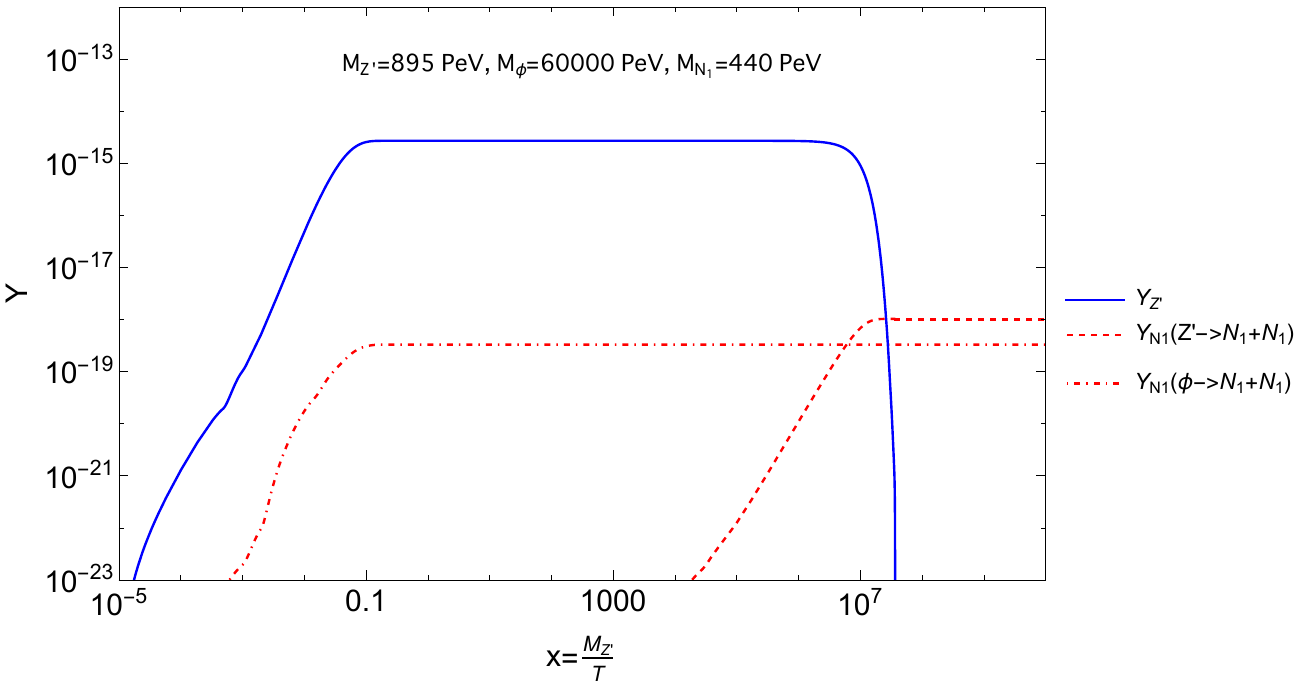}
    \caption{Evolution of $Z'$ and DM $N_{1}$ as a function of $x=\frac{M_{Z'}}{T}$ for a benchmark point.}
    \label{fig:DM_evolution}
\end{figure}

Finally, we show the viability of generating the observed relic of RHN DM of mass 440 PeV in this model by solving the coupled Boltzmann equations in Eq. \eqref{dmeq}. Fig. \ref{fig:DM_evolution} shows the non-thermal production of $Z'$ and DM $N_{1}$ as a function of $x=M_{Z'}/T$ for one particular benchmark point corresponding to $M_{Z'} = 895$ PeV, $M_{\phi} = 60000$ PeV, and DM mass $M_{N_{1}} = 440$ PeV. We consider the $B-L$ symmetry breaking scale to be $v_{\rm BL}=4\times 10^{19}$ GeV which implies $g_{\rm BL}\sim  10^{-11}$ and $Y_{N_{11}} \sim 8\times 10^{-12}$. For this particular benchmark point, the production of DM from non-thermal $Z'$ dominates over the production of DM from thermal $\phi$. The asymptotic comoving abundance of $N_1$ gives rise to present DM relic $\Omega h^2=0.12$ consistent with PLANCK data \cite{Planck:2018vyg}. The required lifetime of DM can be obtained by appropriate choice of Yukawa coupling $Y^{\alpha 1}_D$ without affecting DM relic. We can also produce DM non-thermally from scattering of SM bath particles into $N_1$ mediated by $Z'$, applicable when $Z' \rightarrow N_1 N_1$ is kinematically forbidden. As shown in \cite{Okada:2020cue}, we can get correct DM relic in such a case for $v_{\rm BL} \gtrsim 10^{13}$ GeV allowing sub-Planckian symmetry breaking scales.

We have chosen a minimal DM scenario which contains a long-lived RHN DM, produced non-thermally in the early Universe via scalar or $Z'$ decays. DM mass, DM lifetime as well as $U(1)_{B-L}$ breaking scale are chosen in a way to fit the KM3-230213A event energy and flux. Such new physics scales at $\mathcal{O}(100)$ PeV can, however, be motivated in a variety of beyond standard model frameworks. For example, such a scale can arise as an intermediate scale in non-supersymmetric $SO(10)$ grand unified theories \cite{Bertolini:2009qj}. The framework discussed in \cite{Boyle:2018tzc, Boyle:2018rgh} has a stable RHN DM whose relic matches with observed DM relic only for mass $\sim 480$ PeV. On the other hand, if RHN DM is stabilised by a global symmetry like $Z_2$ forbidding its coupling to lepton and Higgs, Planck scale suppressed operators like $(H^\dagger H)\overline{l_{L}}\tilde{H} N_{R}/M^2_{\rm Pl}$ violating $Z_2$ explicitly while being gauge invariant can lead to DM decay. Such operators arise naturally given the fact that any theory of quantum gravity is expected to violate global symmetries explicitly \cite{Kallosh:1995hi, Witten:2017hdv, Rai:1992xw}. This can lead to an effective Yukawa coupling $Y_{\rm eff} \sim 10^{-32}$ (for $v\sim 246$ GeV, $M_{\rm Pl} =2.4\times 10^{18}$ GeV) keeping it in the required ballpark as given in Eq. \eqref{lifetimedm}.

\section{Other dark matter scenarios}
\label{sec3a}
Since RHN DM decays into neutrino and other SM particles via two-body decay with the same Yukawa coupling controlling all these partial decay widths, it is difficult to reduce the gamma-ray flux without reducing the neutrino flux. This is observed while comparing fluxes for two different $\tau_{\rm DM}$ values in Fig. \ref{fig1} discussed in the previous section. One can make suitable changes to this model in order to decouple the gamma-ray and neutrino flux to some extent. Here we briefly outline two such possibilities.
\subsection{Singlet Scalar DM}
We can consider another complex singlet scalar $\Phi'$ added to the model such that $M_{\phi'} < M_i$. It couples to RHNs as $y_{\phi'} \Phi' \overline{N^c_i}N_i$. Unlike $\Phi$, this new singlet scalar does not acquire VEV and be a long-lived DM. Since RHNs mix with active neutrinos after electroweak symmetry breaking we can have scalar DM decay as $\Phi' \rightarrow \nu \nu$ \cite{Mohapatra:2023aei}. In such a case DM decay to charged particles will arise only at three-body or higher final states, possibly giving a softer spectra for diffuse gamma-rays. Such charged final states arise via $\phi \rightarrow \nu \ell^{\pm} W^{\mp}$. One can also have production of other SM particles via similar three-body decay processes. Such a scalar singlet DM can be generated non-thermally via scalar portal or $Z'$ portal interactions, similar to RHN DM discussed above.

\subsection{Doublet Scalar DM}
Instead of considering Majorana neutrino, one can consider RHNs to be the right chiral parts $\nu_R$ of sub-eV Dirac neutrinos. In this simplest version, they can acquire sub-eV neutrino masses from the SM Higgs itself with tiny Yukawa couplings. In such a case, the gauged $B-L$ symmetry needs to be broken by a singlet scalar with $B-L$ charge different from 2 such that Majorana mass terms are not dynamically generated.

Let us consider a neutrinophilic scalar doublet $\eta$ which couples to neutrinos as $y_\eta \overline{l}_L \tilde{\eta} \nu_R$, where $\nu_R$ is the right chiral part of light Dirac neutrinos. Assuming a $Z_2$ symmetry in the scalar sector under which $\eta \rightarrow -\eta$, the scalar potential is similar to the inert doublet model given by
\begin{align}
    V  & =\mu_H^2|H|^2+\mu_\eta^2|\eta|^2+ \lambda_1|H|^4 + \lambda_2|\eta|^4 +\lambda_3 |H|^2|\eta|^2 \nonumber \\
    & +\lambda_4 |\eta^\dagger H|^2 +\lambda_5[(\eta^\dagger H)^2 +{\rm h.c.}].
\end{align}
After electroweak symmetry breaking, the components of $\eta$ split and the lightest neutral real scalar $\eta_R$ can be DM. Since the scalar sector has a $Z_2$ symmetry and $\mu^2_\eta >0$, the neutral component of $\eta$ does not acquire any VEV. Therefore, DM in this setup can decay only via the Yukawa interactions which break $Z_2$ symmetry. We consider the corresponding Yukawa coupling $y_\eta$ to be tuned such that DM is still long-lived on cosmological scales. Its decay $\eta_R \rightarrow \nu_L \nu_R$ can give rise to the observed neutrino event while its decay into charged final states appear only via three-body decay like $\eta_R \rightarrow \ell^{\pm} W^{\mp} \nu_R$ keeping the resulting secondary gamma-ray spectrum softer compared to the RHN DM studied earlier. While pure thermal scalar doublet DM will be overproduced for DM mass above the unitarity limit, incomplete thermalisation with suitable reheating temperature can generate the desired relic \cite{Okada:2021uqk}.

\section{Conclusion}
\label{sec4}
We have explored the possibility of generating the recently detected highest energy neutrino event of 220 PeV energy by the KM3NeT collaboration from decay of superheavy dark matter. Considering the simplest possibility of a right handed neutrino coupling with Yukawa coupling to leptons in type-I seesaw model to be the DM candidate, we calculate the flux of neutrino and gamma-ray by considering all possible two-body decay modes of DM. Considering the best-fit value of the neutrino flux reported by the KM3NeT collaboration, we find that the present gamma-ray constraints disfavour the DM decay origin of the event, in addition to the mild tension with non-observation of similar events at IceCube and PAO. Due to the large uncertainty associated with the KM3-230213A flux at present, it is still possible to explain the event at within $3\sigma$ CL while being consistent with the above-mentioned constraints. As we consider two-body decay of DM, the preferred DM mass is 440 PeV, the relic of which can be generated non-thermally, if we embed the type-I seesaw scenario in a gauged $U(1)_{B-L}$ setup where additional scalar and vector boson play the crucial role in DM production. We also outline the possibility of two more DM realisations where gamma-ray constraints can be made weaker. To conclude, the possibility of DM origin of the KM3-230213A event is marginal at present due to gamma-ray constraints and non observation of similar events at other experiments. More precise estimates of flux, multi-messenger observations or identification of astrophysical sources in future will be able to settle this hypothesis more concretely.

\acknowledgements
We thank P. S. Bhupal Dev for useful comments. The work of D.B. is supported by the Science and Engineering Research Board (SERB), Government of India grants MTR/2022/000575 and CRG/2022/000603. D.B. also acknowledges the support from the Fulbright-Nehru Academic and Professional Excellence Award 2024-25. D.B. is thankful to the Department of Physics and Astronomy, University of Alabama, Tuscaloosa for kind hospitality during February 11-13, 2025 when this work was initiated.  The work of N.D. is supported by the Ministry of Education, Government of India via the Prime Minister's Research Fellowship (PMRF) December 2021 scheme. The work of N.O. is supported in part by the United States Department of Energy Grant Nos. DE-SC0012447 and DE-SC0023713.

\providecommand{\href}[2]{#2}\begingroup\raggedright\endgroup

\end{document}